\documentclass[review]{elsarticle}

\usepackage{caption}
\usepackage{subcaption}
\usepackage{amsmath}
\usepackage{amssymb}
\usepackage[numbers]{natbib}
\usepackage[version=3]{mhchem}
\usepackage{multirow}
\usepackage{xr}
\usepackage{xcolor}
\usepackage[normalem]{ulem}
\usepackage{mciteplus}
\mciteErrorOnUnknownfalse

\usepackage{float}
\usepackage{bm}
\usepackage{longtable}
\usepackage{multirow}
\usepackage{array}

\newcolumntype{P}[1]{>{\centering\arraybackslash}p{#1}}
\newcolumntype{M}[1]{>{\centering\arraybackslash}m{#1}}
\newcommand{\argmax}{\mathop{\mathrm{argmax}}}
\newcommand{\argmin}{\mathop{\mathrm{argmin}}}

\usepackage{lineno}
\modulolinenumbers[5]

\journal{Materials Today}

\begin{document}

\begin{frontmatter}

\title{Accelerating Materials Discovery with Bayesian Optimization and Graph Deep Learning}

\author[address]{Yunxing Zuo\corref{firstauthor}}
\author[address]{Mingde Qin\corref{firstauthor}}
\author[address]{Chi Chen}
\author[address]{Weike Ye}
\author[address]{Xiangguo Li}
\author[address]{Jian Luo\corref{correspondingauthor}}
\author[address]{Shyue Ping Ong\corref{correspondingauthor}}

\cortext[firstauthor]{These authors contribute equally to this work}
\cortext[correspondingauthor]{Correspondences: jluo@ucsd.edu, ongsp@eng.ucsd.edu}
\address[address]{Department of NanoEngineering, University of California San Diego, 9500 Gilman Dr, Mail Code 0448, La Jolla, CA 92093-0448, United States}

\begin{abstract}
Machine learning (ML) models utilizing structure-based features provide an efficient means for accurate property predictions across diverse chemical spaces. However, obtaining equilibrium crystal structures typically requires expensive density functional theory (DFT) calculations, which limits ML-based exploration to either known crystals or a small number of hypothetical crystals. Here, we demonstrate that the application of Bayesian optimization with symmetry constraints using a graph deep learning energy model can be used to perform ``DFT-free'' relaxations of crystal structures. Using this approach to significantly improve the accuracy of ML-predicted formation energies and elastic moduli of hypothetical crystals, two novel ultra-incompressible hard materials \ce{MoWC2} ($P6_{3}/mmc$) and \ce{ReWB} ($Pca2_{1}$) were identified and successfully synthesized via \textit{in-situ} reactive spark plasma sintering from a screening of 399,960 transition metal borides and carbides. This work addresses a critical bottleneck to accurate property predictions for hypothetical materials, paving the way to ML-accelerated discovery of new materials with exceptional properties.
\end{abstract}


\end{frontmatter}

\section{Introduction}

The accurate prediction of novel stable crystals and their properties is a fundamental goal in computation-guided materials discovery. While \textit{ab initio} approaches such as density functional theory (DFT) \cite{kohnSelfConsistentEquationsIncluding1965, hohenbergInhomogeneousElectronGas1964} have been phenomenally successful in this regard \cite{greeleyComputationalHighthroughputScreening2006, gautierPredictionAcceleratedLaboratory2015, yanSolarFuelsPhotoanode2017, wangMiningUnexploredChemistries2018}, their high computational cost and poor scalability have limited the broad application across vast chemical and structural spaces. As a result, high-throughput DFT screening has been mostly performed on $\sim O(100-1000)$ crystals with relatively small unit cells. 

To circumvent this limitation, machine learning (ML) has emerged as a new paradigm for developing efficient surrogate models for predicting materials properties at scale \cite{sekoPredictionLowThermalConductivityCompounds2015, xueAcceleratedSearchMaterials2016, faberMachineLearningEnergies2016,  sanchez-lengelingInverseMolecularDesign2018, schmidtRecentAdvancesApplications2019, kimInverseDesignPorous2020}. Such ML models are usually trained on large databases of materials properties \cite{jainCommentaryMaterialsProject2013, saalMaterialsDesignDiscovery2013, curtaroloAFLOWLIBORGDistributed2012} to learn the relationship between input chemical and/or structural features and target properties (e.g., formation energies, band gaps, elastic moduli, etc.). Only ML models utilizing structural as well as chemical features can distinguish between polymorphs and be universally applied in materials discovery across diverse crystal structures. In particular, graph neural networks, where atoms and bonds in crystals are represented as nodes and edges in a mathematical graph, have emerged as a particularly promising approach with state-of-the-art accuracy in predicting a broad range of energetic, electronic and mechanical properties \cite{chenGraphNetworksUniversal2019, chenLearningPropertiesOrdered2021, xieCrystalGraphConvolutional2018, duvenaudConvolutionalNetworksGraphs, kearnesMolecularGraphConvolutions2016, gilmerNeuralMessagePassing2017, schuttSchNetDeepLearning2018, jorgensenNeuralMessagePassing2018}.

Ironically, a critical bottleneck in the application of structure-based ML models for materials discovery is the requirement for equilibrium crystal structures as the inputs. These are obtained by ``relaxing'' initial input structures along their potential energy surfaces, which are typically computed via expensive DFT calculations. While there have been recent efforts \cite{schuttSchNetPackDeepLearning2019, cheonCrystalStructureSearch2020} at deriving accurate interatomic forces from graph representations, the application has been limited to a few molecular systems or constrained chemical spaces.

Here, we propose a Bayesian Optimization With Symmetry Relaxation (BOWSR) algorithm to obtain equilibrium crystal structures for accurate ML property predictions without DFT. Utilizing a highly efficient MatErials Graph Network (MEGNet) formation energy model \cite{chenGraphNetworksUniversal2019}, we demonstrate that BOWSR-relaxed structures can serve as accurate inputs to ML property models, yielding far higher accuracy in the prediction of various materials properties. Finally, we demonstrate the power of this approach by screening $\sim$400,000 transition metal borides and carbides for ultra-incompressible hard materials. Two new materials with relatively low predicted energy above hull \cite{ongLiFePhase2008} were successfully synthesized and demonstrated to have exceptional mechanical properties, in line with the ML prediction.

\section{Results}

\subsection{Bayesian Optimization With Symmetry Relaxation Algorithm}

\begin{figure}
    \centering
    \includegraphics[width=0.8\textwidth]{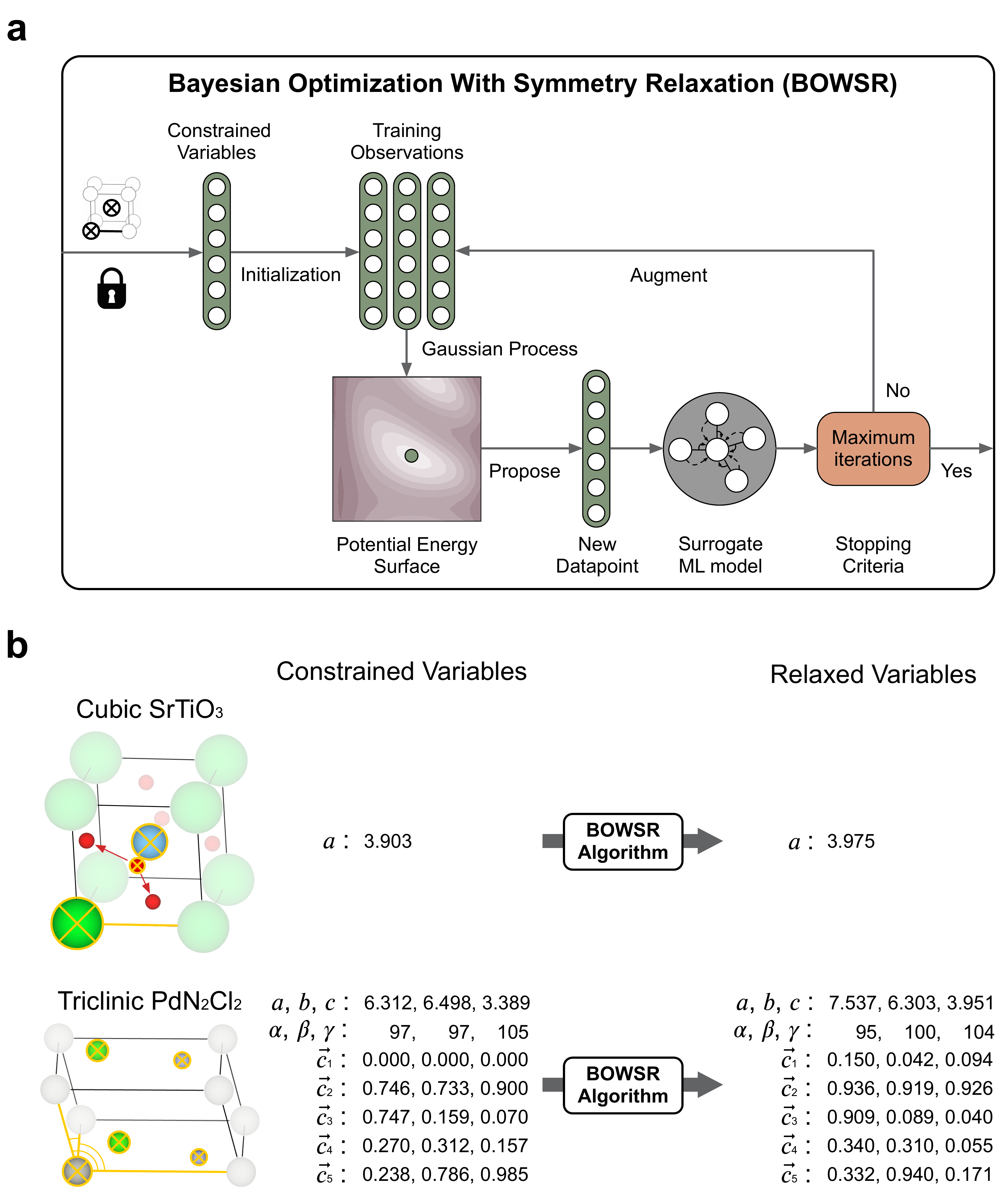}
    \caption{\textbf{Bayesian Optimization With Symmetry Relaxation (BOWSR) algorithm.} \textbf{a,} The BOWSR algorithm parameterizes each crystal based on the independent lattice parameters and atomic coordinates based on its space group. The potential energy surface is then approximated by initializing a set of training observations and energies from the ML energy model. Bayesian optimization is then used to iteratively propose lower energy geometries based on prior observations. \textbf{b, } Two examples of the geometry parameterization for cubic perovskite \ce{SrTiO3} and triclinic \ce{PdN2Cl2}. For the high-symmetry cubic perovskite, the lattice parameter $a$ is the only independent parameter, and all atoms are in special Wyckoff positions with no degrees of freedom in the fractional coordinates. For the triclinic crystal, all six lattice parameters and all atomic coordinates are independent parameters.}
    \label{fig:bowsralgo}
\end{figure}

Bayesian optimization (BO) is an adaptive strategy for the global optimization of a function. In crystal structure relaxation, the function to be optimized is the potential energy surface, which expresses the energy of the crystal as a function of the lattice parameters and atomic coordinates. In the BOWSR algorithm, the symmetry (space group) of the lattice and the Wyckoff positions of the atoms are constrained during the relaxation process, i.e., only the independent lattice parameters and atomic coordinates are allowed to vary. The BO goal is then the minimization of the following mapping:
    \begin{equation}
        x:=\{a, b, c, \alpha, \beta, \gamma, \vec{c_1}, \vec{c_2}, ...\}
        \label{eqn:independentparameters}
    \end{equation}
    \begin{equation}
        x_{\rm opt} = \argmin_x U(x), U: R^n  \xrightarrow{} R
        \label{eqn:optimization}
    \end{equation}
where \{$a, b, c, \alpha, \beta, \gamma$\} and \{$\vec{c_1}, \vec{c_2}, ...$\} represent the independent lattice parameters and the atomic positions for a $P1$ crystal, respectively, and $U(.)$ is the energy of the system. The schematic of the BOWSR algorithm as well as two examples of the geometry parameterization for a high-symmetry cubic crystal and a low-symmetry triclinic crystal are shown in Figure \ref{fig:bowsralgo}.

The convergence accuracy and speed of BOWSR are set by the energy evaluator $U(.)$, which can be any computational model, including \textit{ab initio} methods, empirical potentials, and surrogate ML models. In this work, we have elected to use a MatErials Graph Network (MEGNet) formation energy model previously trained on the DFT-computed formation energies of 133,420 Materials Project crystals \cite{jainCommentaryMaterialsProject2013}. This MEGNet model has a cross-validated mean absolute error (MAE) of 26 meV atom$^{-1}$, which is among the best accuracy among general ML models thus far \cite{schuttSchNetDeepLearning2018,xieCrystalGraphConvolutional2018}. Examples of the convergence of the BOWSR algorithm using the MEGNet energy model are shown in Figure S1.

\subsection{Properties Prediction}

Elemental substitution is a common, chemically intuitive approach to deriving potential new compounds. For instance, the rock salt \ce{LiCl} can be derived from rock salt \ce{NaCl} by substituting Na for the chemically similar Li. Here, we demonstrate the potential for BOWSR to substantially enhance ML property predictions of the formation energies and elastic moduli (bulk and shear moduli) of substituted crystals. The dataset comprises a total of 12,277 and 1,672 unique crystals with pre-computed DFT formation energies and elastic moduli, respectively, from the Materials Project \cite{jainCommentaryMaterialsProject2013}. These crystals belong to 144 (35 binary, 91 ternary, and 18 quaternary) common structure prototypes in the Inorganic Crystal Structure Database (ICSD) \cite{bergerhoffInorganicCrystalStructure1983, belskyNewDevelopmentsInorganic2002}. Each prototype comprises at least 30 compositions (statistical distribution shown in Figure S2). For each crystal in the dataset (e.g., rock salt GeTe), another crystal with the same prototype but a different composition (e.g., rock salt NaCl) was selected at random and multi-element substitutions (Na$\rightarrow$Ge, Cl$\rightarrow$Te) were performed to arrive at an ``unrelaxed'' structure. The BOWSR algorithm was then applied to obtain a BOWSR-relaxed structure. The unrelaxed, BOWSR-relaxed, and the original DFT-relaxed structures were then used as inputs for property predictions using MEGNet models. These MEGNet models were trained on the DFT-computed formation energies and elastic moduli of 133,420 and 12,179 crystals, respectively, from the Materials Project.

\begin{figure}[H]
    \centering
    \includegraphics[width=1\textwidth]{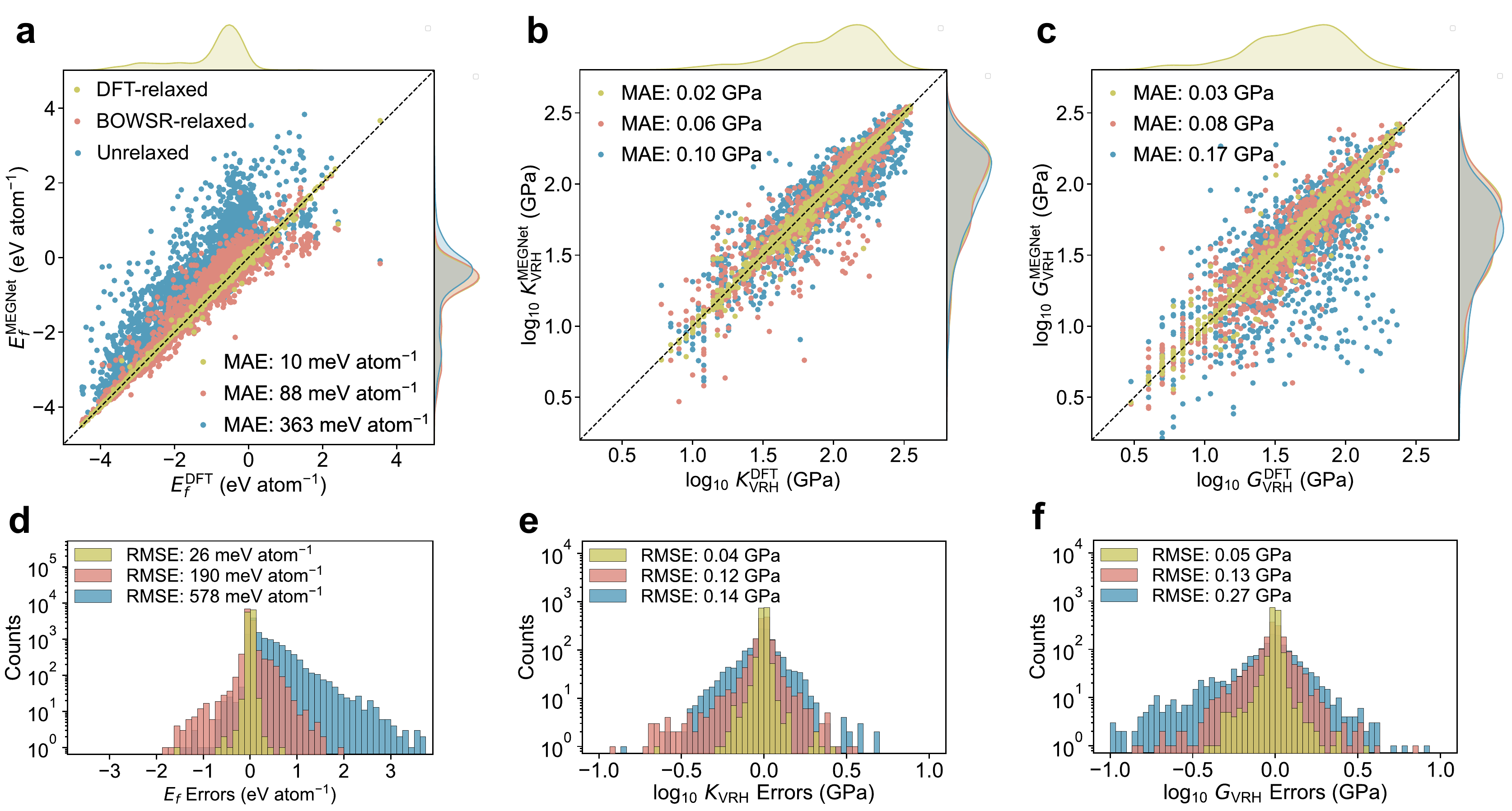}
    \caption{\textbf{Performance of MEGNet materials properties predictions for three levels of relaxation.} Three levels of relaxation are served as inputs for MEGNet materials properties predictions --- unrelaxed, BOWSR-relaxed, and DFT-relaxed structures. \textbf{a,} \textbf{b,} and \textbf{c } show the parity plot of MEGNet prediction on formation energies, bulk moduli, and shear moduli, respectively. \textbf{d,} \textbf{e,} and \textbf{f} show the errors distribution on the corresponding materials property.}
    \label{fig:bowsrmae}
\end{figure}

Figure \ref{fig:bowsrmae} compares the MEGNet model predictions using the unrelaxed, BOWSR-relaxed, and DFT-relaxed structures as inputs with respect to DFT-computed values. The mean absolute errors (MAEs) of the MEGNet models using the DFT-relaxed structures provide a best-case performance baseline. It should be noted that the MEGNet models were trained using a superset of data from the Materials Project that includes the DFT-relaxed structures \cite{jainCommentaryMaterialsProject2013}. Hence, the reported MAEs of MEGNet predictions using DFT-relaxed structures in this work are much smaller than the previously reported MAEs of these MEGNet models from cross-validation and should not be considered as a metric for MEGNet performance. Using the unrelaxed structures as inputs results in much higher, positively skewed MAEs in the MEGNet formation energy prediction compared to using DFT-relaxed structures. This is because the unrelaxed structures have lattice parameters and atomic positions that can deviate substantially from the ground state DFT-relaxed structures, resulting in higher energies. Using the BOWSR-relaxed structures as inputs reduces the MAEs by a factor of four, from 363 meV atom$^{-1}$ to 88 meV atom$^{-1}$. The $R^2$ also substantially increases from 0.67 to 0.96, and the error distribution is Gaussian-like with a mean close to 0. Similarly, large improvements in the MEGNet predictions of the elastic moduli are also observed using the BOWSR-relaxed structures compared to using unrelaxed structures, with MAEs in the $\log_{10} K_{\rm VRH}$ and $\log_{10} G_{\rm VRH}$ reducing by half. We tested the sensitivity of the BOWSR algorithm to the initial structures used to perform elemental substitution. Using four randomly chosen parent structures with different lattice parameters for each prototype, the above procedures were repeated and the results are shown in Figure S3. The BOWSR-relaxed structures exhibit consistently low errors regardless of initial structure selection.

We also tested the sensitivity of the BOWSR algorithm to the accuracy of the energy evaluator by artificially introducing Gaussian noise into the MEGNet formation energy prediction. The energy errors from the BOWSR-relaxed structures are linearly correlated with the errors of the surrogate ML model with the root mean square error (RMSE) ranging from 27 to 1000 meV $\rm atom^{-1}$ (Figure S4), which indicates the robustness of the BO propagation and the broad applicability of the BOWSR algorithm to any surrogate ML models. The same linear correlations are also observed between the elastic moduli errors and the errors of the surrogate ML model, as shown in Figure S4c and S4d.

\subsection{Discovery of Ultra-incompressible Hard Materials}

\begin{figure}[H]
    \centering
    \includegraphics[width=0.5\textwidth]{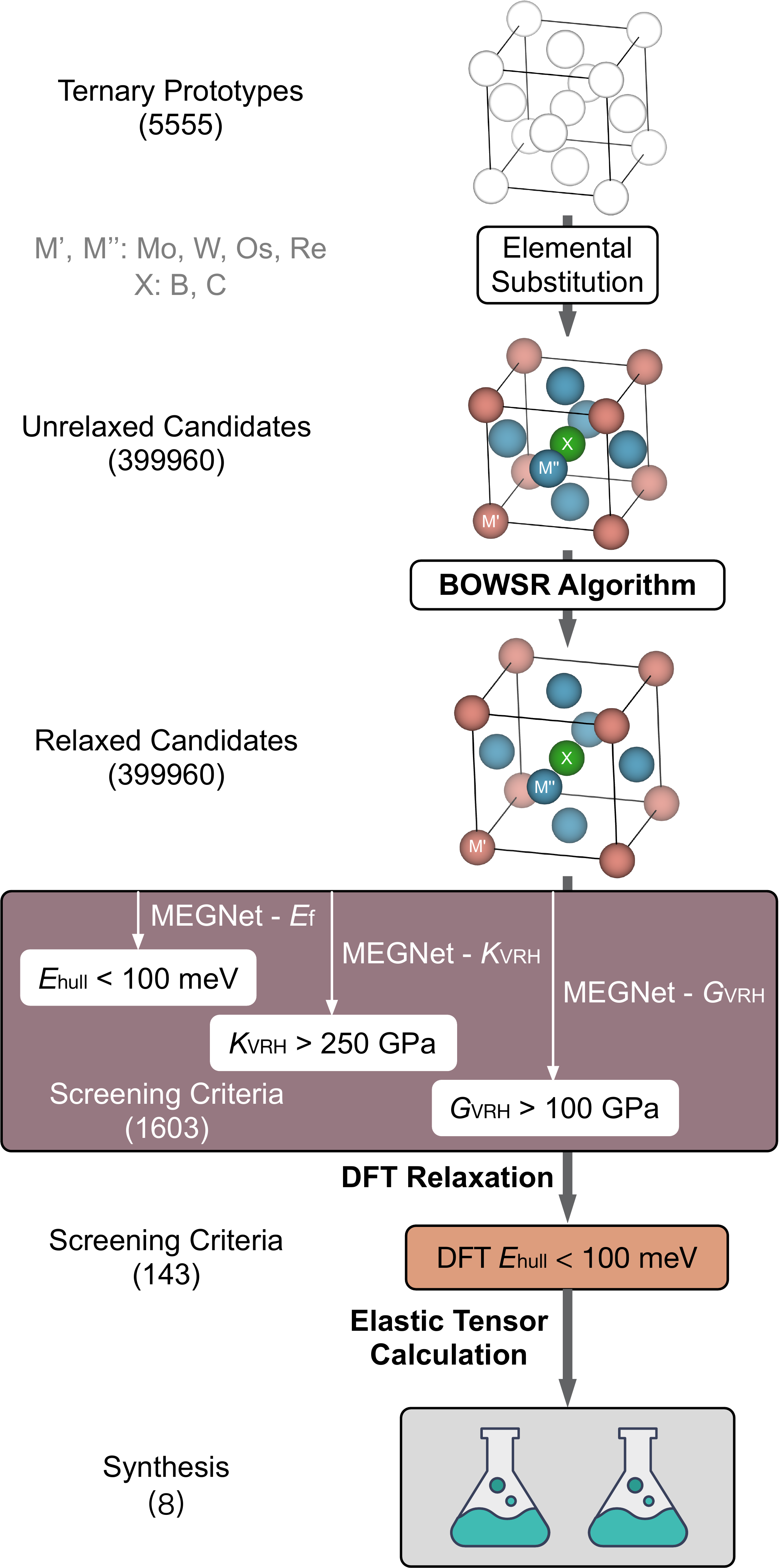}
    \caption{\textbf{Flowchart of ultra-incompressible materials discovery leveraged by the BOWSR algorithm and MEGNet models.} The materials candidates were generated by elemental substitution to structure prototypes. These candidates were relaxed by the BOWSR algorithm and subsequently screened by their predicted thermodynamic stability and mechanical properties. The screened candidates were further verified by DFT calculations, and the high-ranked candidates were directed to synthesis attempts. Quantity in the parenthesis shows the number of candidates at each stage.}
    \label{fig:discovery_workflow}
\end{figure}

We used the BOWSR algorithm with the MEGNet models to rapidly screen hundreds of thousands of candidates for novel ultra-incompressible hard materials, as shown in Figure \ref{fig:discovery_workflow}. Given that binary compounds have already been extensively explored in the literature \cite{akopovRediscoveringCrystalChemistry2017, yeungUltraincompressibleSuperhardMaterials2016}, we targeted our search in 12 ternary M$'_x$M$''_y$X$_z$ chemical spaces, where M$'$, M$''$ = Mo, W, Os, or Re and X = B or C. These elements were selected based on their common occurrences in ultra-incompressible hard binary compounds. By combinatorially applying elemental substitutions to 5,555 ternary structures prototypes in the ICSD \cite{bergerhoffInorganicCrystalStructure1983, belskyNewDevelopmentsInorganic2002}, 399,960 candidates were generated and relaxed using the BOWSR algorithm with the MEGNet energy model. The BOWSR-relaxed candidates were then screened for stability and mechanical properties using MEGNet property models. The stability metric used was the energy above hull $E_{\rm hull}^{\rm MEGNet}$, which was computed using the predicted formation energy $E_{\rm f}^{\rm MEGNet}$ with the 0 K phase diagram in the Materials Project database \cite{ongLiFePhase2008, jainCommentaryMaterialsProject2013, yeDeepNeuralNetworks2018}. At this intermediate stage, a relatively generous threshold of $E_{\rm hull}^{\rm MEGNet} < 100$ meV atom$^{-1}$ was used to obtain candidates that are likely to be thermodynamic stable \cite{sunThermodynamicScaleInorganic2016}. Of these, candidates with relatively high MEGNet-predicted bulk and shear moduli ($K_{\rm VRH}^{\rm MEGNet} > 250$ GPa and $G_{\rm VRH}^{\rm MEGNet} > 100$ GPa) were identified. Similar to the stability criterion, the mechanical criteria used are slightly lower than the conventional threshold for ultra-incompressibility to account for the higher MAE of the MEGNet elastic moduli predictions \cite{yeungUltraincompressibleSuperhardMaterials2016}. DFT relaxations and energy calculations were then carried out on the 1,603 candidates that passed all three ML-based screening criteria. Subsequently, expensive DFT elastic tensor calculations \cite{dejongChartingCompleteElastic2015} were performed on the 143 candidates that have DFT $E_{\rm hull}^{\rm DFT} < 100$ meV atom$^{-1}$.

\begin{figure}[H]
    \centering
    \includegraphics[width=1\textwidth]{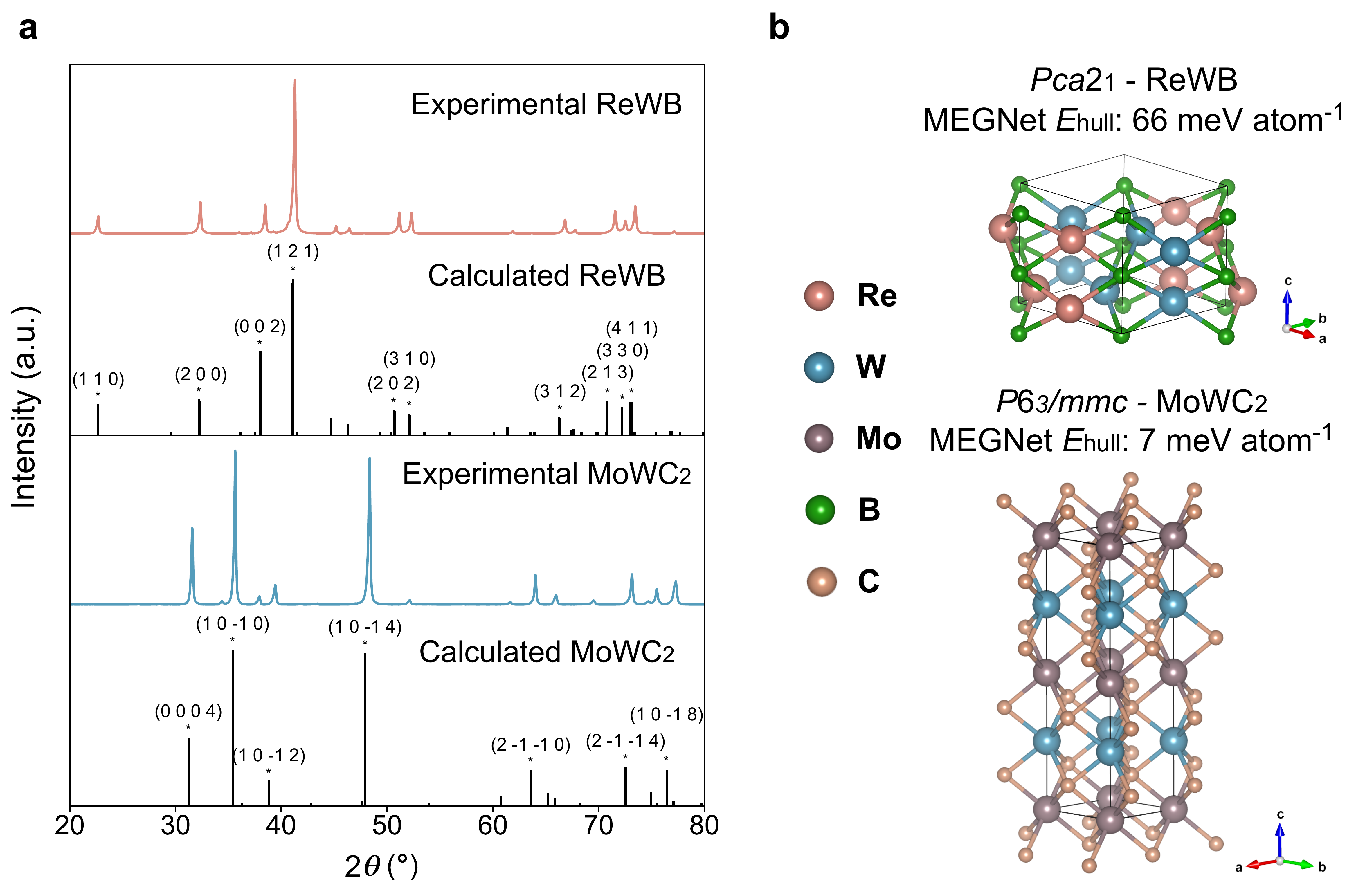}
    \caption{\textbf{Two new materials proposed by the BOWSR algorithm and MEGNet prediction confirmed by XRD characterizations.} \textbf{a,} Measured and calculated XRD patterns of two materials (\ce{ReWB} and \ce{MoWC2}). The major peaks are indexed for reference. The pymatgen library \cite{ongPythonMaterialsGenomics2013} was used to calculate the XRD patterns of the DFT-relaxed crystal structures. Minor shifts in peak positions between the measured and calculated XRD patterns can be attributed to the systematic errors between DFT and experimentally-measured lattice parameters. The comparison in lattice parameters between BOWSR-relaxed, DFT-relaxed, and experimental structures is shown in Table S1. \textbf{b,} Crystal structures and space group of these two materials. The predicted energy above hull for \ce{ReWB} and \ce{MoWC2} are 66 and 7 meV atom$^{-1}$, respectively.}
    \label{fig:xrd}
\end{figure}

Table \ref{table:bulk_modulus} summarizes the computed elastic properties of the top ten candidates with the highest computed bulk modulus together with other well-known ultra-incompressible materials. Attempts were then made to synthesize all ten candidates with eight unique compositions via \textit{in-situ} reactive spark plasma sintering (SPS) using elemental precursors in the appropriate ratios (see Methods). Two crystals, \ce{MoWC2} ($P6_{3}/mmc$) and \ce{ReWB} ($Pca2_{1}$), were successfully synthesized and confirmed via X-ray diffraction (XRD, Figure \ref{fig:xrd}a) as single phase, while the synthesis of the other six compositions yielded multiple phases (see Figure S5 - S10). Henceforth, we will refer to the two novel phases of \ce{MoWC2} ($P6_{3}/mmc$) and \ce{ReWB} ($Pca2_{1}$) simply as \ce{MoWC2} and \ce{ReWB}, respectively.

\begin{table}[htbp]
  \centering
  \caption{\textbf{DFT-computed bulk modulus ($\boldsymbol{K}_{\rm \textbf{VRH}}$), shear modulus ($\boldsymbol{G}_{\rm \textbf{VRH}}$), Young's modulus ($\boldsymbol{E}_{\rm \textbf{VRH}}$), Poisson's ratio ($\boldsymbol{\nu}$) and energy above hull ($\boldsymbol{E}_{\rm \textbf{hull}}$) for the top 10 candidates with regard to $\boldsymbol{K}_{\rm \textbf{VRH}}$ in descending order}. \ce{MoWC2} and \ce{ReWB} are bolded as they are successfully synthesized by experiments. Some of the known ultra-incompressible materials are used as references.}
  \makebox[\linewidth]{
  \begin{tabular}{c|ccccc}\hline
    & $K_{\rm VRH}$ (GPa) & $G_{\rm VRH}$ (GPa) & $E_{\rm VRH}$ (GPa) & $\nu$ &     $E_{\rm hull}$ (meV atom$^{-1}$) \\\hline
    Candidates &  &  &  &  \\
    \hline
    \ce{ReOsB} ($P\bar{6}m2$)       & 370.7 & 241.3 & 594.7 & 0.233 & 31.7 \\
    \hline
    \ce{ReOsB2} ($P6_{3}/mmc$)       & 367.3 & 220.9 & 552.0 & 0.250 & 87.8 \\
    \hline
    \textbf{\ce{MoWC2}} ($\textbf{\textit{P}}\textbf{6}_{\textbf{3}}/\textbf{\textit{mmc}}$)    & \textbf{357.9} & \textbf{260.5} & \textbf{628.8} & \textbf{0.207} & \textbf{96.3} \\
    \hline
    \ce{ReWB} ($Fddd$)      & 356.8 & 176.9 & 455.5 & 0.287 & 20.6  \\
    \hline
    \ce{Re13WB9} ($P\bar{6}m2$)      & 353.1 & 177.0 & 455.1 & 0.285 & 88.4 \\
    \hline
    \textbf{\ce{ReWB}} ($\textbf{\textit{Pca}}\textbf{2}_{\textbf{1}}$)      & \textbf{352.6} & \textbf{144.1} & \textbf{380.4} & \textbf{0.320} & \textbf{33.1}\\
    \hline
    \ce{OsWB} ($Pbam$)      & 351.1 & 183.1 & 467.9 & 0.278 & 43.3  \\
    \hline
    \ce{ReWB} ($Cmce$)      & 350.9 & 161.5 & 420.1 & 0.301 & 32.6 \\
    \hline
    \ce{Re6W7B8} ($P6/m$)      & 348.4 & 182.8 & 466.8 & 0.277 & 22.2 \\
    \hline
    \ce{ReW2B2} ($P4/mbm$)      & 345.8 & 156.0 & 406.8 & 0.304 & 72.1\\
    \hline
    Known materials &  &  &  &  \\
    \hline
    \ce{C} ($Fd\bar{3}m$) & 430.3 & 503.6 & 1086.9 & 0.079 & 136.4 \\
    \hline
    \ce{WC} ($P\bar{6}m2$) & 389.8 & 280.0 & 677.8 & 0.210 & 1.1 \\
    \hline
    \ce{BN} ($F\bar{4}3m$)    & 370.1 & 382.8 & 852.4 & 0.116 & 77.3 \\
    \hline
    \ce{ReB2} ($P6_{3}/mmc$) & 334.9 & 272.3 & 642.7 & 0.180 & 4.7 \\
    \hline
    \end{tabular}}
    \label{table:bulk_modulus}
\end{table}%

\begin{figure}[H]
    \centering
    \includegraphics[width=1\textwidth]{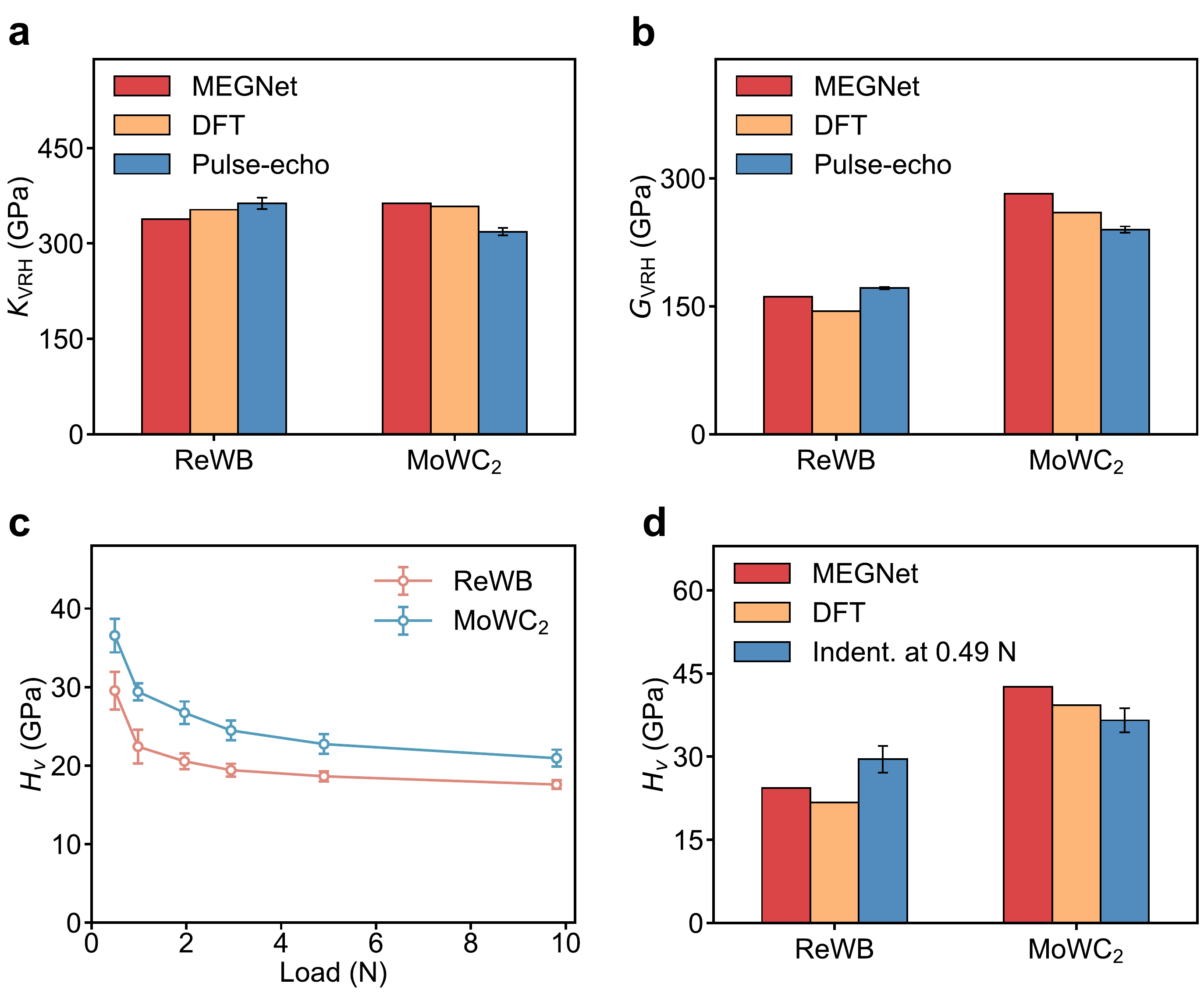}
    \caption{\textbf{Experimental measurements and theoretical prediction of mechanical properties for the new materials.} Comparisons of \textbf{a, } bulk moduli and \textbf{b, } shear moduli between MEGNet, DFT and pulse-echo measurements. \textbf{c,} Measured Vickers microhardness of \ce{ReWB} and \ce{MoWC2} under loads ranging from 0.49 N to 9.8 N. \textbf{d,} Comparisons between the hardness derived from MEGNet predicted bulk modulus and shear modulus \cite{chenIntrinsicCorrelationHardness2011} and the hardness measured by Vickers indentation at the low load (0.49 N). The DFT calculated results are referenced. The error bar represents the standard deviation of multiple experimental measurements.}
    \label{fig:elastic_hardness}
\end{figure}

The mechanical properties of \ce{MoWC2} and ReWB were measured using the pulse-echo method \cite{huntingtonUltrasonicMeasurementsSingle1947, lazarusVariationAdiabaticElastic1949}. As shown in Figure \ref{fig:elastic_hardness}a and \ref{fig:elastic_hardness}b, the experimentally-measured bulk and shear moduli are in excellent agreement with both the MEGNet and DFT predictions. Both new materials exhibit ultra-incompressibility, with bulk modulus close to or larger than 300 GPa \cite{yeungUltraincompressibleSuperhardMaterials2016}. \ce{MoWC2} also exhibits high estimated Vickers hardnesses $H_v$ of 36.6 at 0.49 N indentation load and 20.9 GPa at 9.8 N load (Figure \ref{fig:elastic_hardness}c). ReWB has a comparatively lower measured hardness of 29.5 at 0.49 N load and 17.6 GPa at 9.8 N load. The $H_v$ values at 0.49 N load are within 20-25\% of those derived from the MEGNet and DFT predicted shear moduli via the empirical relation $H_{v}  = 0.151 G$ \cite{chenIntrinsicCorrelationHardness2011}, as shown in Figure \ref{fig:elastic_hardness}d.

\section{Discussion}

Many materials properties, such as formation energies, mechanical properties, etc., exhibit a strong dependence on the crystal structure. However, obtaining equilibrium crystal structures as inputs to accurate ML models still requires expensive \textit{ab initio} computations. By coupling an accurate MEGNet energy model with Bayesian optimization of symmetry-constrained parameters, we demonstrate that the new algorithm can reasonably approximate equilibrium structures. The resulting substantial improvements in ML property predictions enable the rapid screening of $\sim$400,000 candidate crystals for stability and exceptional mechanical properties, $10^{3}$ - $10^{4}$ orders of magnitude larger than that accessible by high-throughput DFT calculations. 

The effectiveness of the BOWSR algorithm is limited by the accuracy of the energy evaluator. While the MEGNet formation energy model has been selected in this work, we foresee that the future development of more accurate ML energy models may improve the quality of the BOWSR-relaxed structures, and subsequent ML property predictions, even further. Here, the search for ultra-incompressible materials has been chosen as a model problem due to the high cost of acquisition of elastic moduli via standard DFT approaches, but the approach outlined can be readily extended to any property for which a reliable ML model can be developed. It should be noted, however, that there is an inverse relationship between the cost of acquisition and the training data size; hence, datasets on expensive properties (e.g., elastic moduli, optical properties, etc.) tend to be much smaller in size compared to cheaper properties, making it more difficult to build reliable ML models for high-cost properties.  While approaches such as transfer learning or multi-fidelity models have been shown to mitigate this trade-off to some extent \cite{chenGraphNetworksUniversal2019, chenLearningPropertiesOrdered2021, pilaniaMultifidelityMachineLearning2017, jhaEnhancingMaterialsProperty2019}, the generally higher errors in ML models for high-cost properties should be factored into the screening process in the thresholds.

\section{Acknowledgements}
This research was primarily supported by the National Science Foundation Materials Research Science and Engineering Center program through the UC Irvine Center for Complex and Active Materials (DMR-2011967). The authors also acknowledge the use of data and software resources from the Materials Project, funded by the U.S. Department of Energy, Office of Science, Office of Basic Energy Sciences, Materials Sciences and Engineering Division under contract no. DE-AC02-05-CH11231: Materials Project program KC23MP, and computing resources provided by the Extreme Science and Engineering Discovery Environment (XSEDE) under grant ACI-1548562.

\section{Material and Methods}

\subsection{Bayesian Optimization With Symmetry Relaxation Algorithm}

Geometry relaxation of a crystal structure requires the optimization of up to $3N+6$ variables - six lattice parameters and three fractional coordinates for each of the $N$ atoms. By constraining the symmetry to remain unchanged during relaxation can reduce the number of independent variables considerably \cite{wangCrystalStructurePrediction2010, glassUSPEXEvolutionaryCrystal2006, lenzParametricallyConstrainedGeometry2019}. The open-source \textit{spglib} \cite{togoTextttSpglibSoftwareLibrary2018} package was used for symmetry determination. The search for optimized symmetry-constrained lattice parameters and atomic coordinates that minimize the total energy was then carried out via Bayesian optimization (BO). The changes in the variables were used as the optimization inputs to reduce the tendency of the BO process being dominated by parameters with large magnitudes. This approach has been previously used for geometry optimization along the imaginary phonon modes \cite{lenzParametricallyConstrainedGeometry2019}.

Using a Latin hypercube sampling, a set of training observations $D\sim\{(x_{i}, U(x_{i}))\ i=1:m\}$ were initialized, where the $x_{i}$ are the $m$ independent lattice parameters and atomic coordinates and $U(.)$ is the energy of the corresponding structure evaluated by the surrogate model (see Equations \ref{eqn:independentparameters} and \ref{eqn:optimization}). The BO strategy comprises two steps \cite{lookmanActiveLearningMaterials2019}:
\begin{enumerate}
    \item A Gaussian process (GP) model is trained on the initialized training observations $D$ to approximate the $U(x)$. The Rational Quadratic kernel \cite{rasmussenGaussianProcessesMachine2006} is adopted as the covariance function of GP. The noise level of GP model is set to the root mean square error (RMSE) of the energy model.
    \item The acquisition function that balances the exploitation and exploration is calculated for samples in the search space apart from the training observations and the candidate with optimal acquisition function is proposed to be evaluated (formation energy prediction by surrogate ML model). Exploitation represents the samples with high predicted mean from the GP, and exploration accounts the samples with high predictive uncertainty \cite{martinez-cantinBayesianExplorationexploitationApproach2009, shahriariTakingHumanOut2016, balachandranAdaptiveStrategiesMaterials2016}. Here, we chose the expected improvement as the acquisition function, which can be analytically expressed as \cite{jonesEfficientGlobalOptimization, lookmanInformationScienceMaterials2016}:
        \begin{equation}\label{expected_improvement}
            E[I(x)] = (\mu(x) - U(x^{+}) - \xi)\cdot\Phi(Z) + \sigma(x)\cdot\phi(Z)
        \end{equation}
    and
        \begin{equation}
            x^{+} = \argmax_{i=1, \dots, n} U(x_{i})
        \end{equation}
        \begin{equation}
            Z = \frac{\mu(x) - U(x^{+}) - \xi}{\sigma(x)}
        \end{equation}
    where $\mu(x)$ and $\sigma(x)$ are the mean and standard deviation of the posterior distribution on $x$ from the GP, respectively, and $\Phi(x)$ and $\phi(x)$ are the cumulative distribution function (CDF) and probability density function (PDF), respectively. The $\xi$ parameter can be tuned to balance the trade-off between the first term (exploitation) and the second term (exploration) in Equation (\ref{expected_improvement}). Until the maximum number of iteration steps is reached, the sample with optimal acquisition function was iteratively augmented to the training observations and used to update the GP surrogate model in the next loop.
\end{enumerate}

\subsection{Materials Graph Network Models}

The Materials Graph Network (MEGNet) models used in this work are based on the same architecture as our previous work \cite{chenGraphNetworksUniversal2019}. Briefly, three graph convolutional layers with [64, 64, 32] neurons were used in each update function, and the shifted softplus function was used as the non-linear activation function. A set2set readout function with two passes was used after the graph convolution steps. The cutoff radius for constructing the neighbor graph was 5 \AA. The MEGNet formation energy ($E_f$) and elasticity ($K_{\mathrm{VRH}}$ and $G_{\mathrm{VRH}}$) models were trained using the 2019.4.1 version of Materials Project database \cite{jainCommentaryMaterialsProject2013} containing 133,420 structure-formation energy and 12,179 structure-bulk/shear modulus data pairs. Each dataset was split into 80\%:10\%:10\% train:validation:test ratios. During the model training, we used a batch size of 128 structures, and set the initial learning rate to 0.001 in the Adam optimizer. All models were trained for a maximum of 1500 epochs with an early stopping callback, which terminates the model training if the validation error does not reduce for 500 consecutive steps. The mean absolute errors (MAEs) of $E_f$, $\log_{10}(K_{\mathrm{VRH}})$ and $\log_{10}(G_{\mathrm{VRH}})$ models in test data are 26 meV atom$^{-1}$, 0.07, and 0.12, respectively.

\subsection{DFT calculations}

The DFT relaxations, energy and elastic tensor calculations for the small number of candidates that passed the ML screening were carried out using Vienna \textit{ab
initio} simulation package (VASP) \cite{kresseEfficientIterativeSchemes1996} within the projector augmented wave approach \cite{blochlProjectorAugmentedwaveMethod1994}. The exchange-correlation interaction was described using the Perdew-Burke-Ernzerhof (PBE) generalized gradient approximation (GGA) \cite{perdewGeneralizedGradientApproximation1996} functional for structural relaxations and energy calculations. The plane wave energy cutoff was set to 520 eV, and the k-point density of at least 1,000 per number of atoms was used. All structures were relaxed with energies and forces converged to $10^{-5}$ eV and 0.01 eV/{\AA}, respectively, consistent with the calculation setting used in the Materials Project \cite{jainCommentaryMaterialsProject2013}. The elastic tensor calculations were performed using the procedure described in previous work \cite{dejongChartingCompleteElastic2015}. A tighter energy convergence criterion of $10^{-7}$ eV was used, and strains with magnitude of (-1\%, -0.5\%, 0.5\%, 1\%) were applied to each of the 6 independent components of strain tensor. 

\subsection{Synthesis}

Bulk specimens of candidates \ce{ReOsB}, \ce{ReOsB2}, \ce{MoWC2}, \ce{ReWB}, \ce{Re13WB9}, \ce{OsWB}, \ce{Re6W7B8}, and \ce{ReW2B2} were synthesized via \textit{in-situ} reactive spark plasma sintering (SPS). Elemental powders of Mo, W ($>$99.5\% purity, $\sim$325 mesh, Alfa Aesar), Re ($\sim$99.99\% purity, $\sim$325 mesh, Strem Chemicals), Os ($\sim$99.8\% purity, $\sim$200 mesh, Alfa Aesar), boron ($\sim$99\% purity, 1-2 $\mu$m, US Research Nanomaterials), and graphite ($\sim$99.9\% purity, 0.4-1.2 $\mu$m, US Research Nanomaterials) were utilized as precursors. For each composition, stoichiometric amounts of elemental powders were weighted out in batches of 5 g. The powders were first mixed by a vortex mixer, and then high energy ball milled (HEBM) in a Spex 8000D mill (SpexCertPrep) by tungsten carbide lined stainless steel jars as well as 11.2 mm tungsten carbide milling media (ball-to-powder ratio $\approx$ 4.5:1) for 50 min. 0.05 g or $\sim$1 wt\% of stearic acid was used as lubricant in the milling process. After HEMB, the as-milled powder mixtures were loaded into 10 mm graphite dies lined with graphite foils in batches of 2.5 g, and subsequently consolidated into dense pellets via SPS in vacuum ($<10^{-2}$ Torr) by a Thermal Technologies 3000 series SPS machine. The HEBM and powder handing were conducted in an argon atmosphere (with \ce{O2} level $<$ 10 ppm) to prevent oxidation.

During the SPS, specimens were initially heated to 1400 $^\circ$C at a rate of 100 $^\circ$C/min under constant pressure of 10 MPa. For the final densification, the temperature was subsequently raised at a constant rate of 30 $^\circ$C/min to a final isothermal sintering temperature, which was set at different levels for different target compositions --- 1800 $^\circ$C (\ce{ReWB}), 1700 $^\circ$C (\ce{MoWC2} and \ce{Re6W7B8}), 1600 $^\circ$C (\ce{Re13WB9} and \ce{ReW2B2}), or 1500 $^\circ$C (\ce{ReOsB}, \ce{ReOsB2}, and \ce{OsWB}), and maintained isothermally for 10 min. Meanwhile, the pressure was increased to 50 MPa at a ramp rate of 5 MPa/min. The final densification temperature was optimized for each specimen to achieve a high relative density while prevent specimen melting due to overheating. The \textit{in-situ} reactions between elemental precursors likely took place during the initial temperature ramping. After sintering, the specimens were cooled down naturally inside the SPS machine (with power off).

\subsection{Experimental Characterization}

Sintered specimens were first ground to remove the carbon-contaminated surface layer from the graphite tooling, and polished for further characterizations. X-ray diffraction (XRD) experiments were conducted using a Rigaku Miniflex diffractometer with the Cu K$\alpha$ radiation at 30 kV and 15 mA. The Vickers microhardness tests were carried out on a LECO diamond microindentor with loading force varying from 0.49 N (50 gf) to 9.8 N (1 kgf) and constant holding time of 15 s, abiding by the ASTM Standard C1327. Over 20 measurements at different locations were conducted for each specimen at each indentation load to ensure statistical validity and minimize the microstructural and grain boundary effects. In particular, over 30 measurements were conducted for each specimen at 9.8 N indentation load.

The Young's and shear moduli of the specimens were calculated from the ultrasonic velocities measured with a Tektronix TDS 420A digital oscilloscope, following the ASTM standard A494-15. Multiple measurements were conducted at different locations.

\section{Data Availability}

The raw/processed data required to reproduce these findings cannot be shared at this time due to legal or ethical reasons.

\section{References}

\bibliography{refs}

\end{document}


\begin{frontmatter}

\title{Supplementary Material\\Accelerating Materials Discovery with Bayesian Optimization and Graph Deep Learning}

\author[address]{Yunxing Zuo\corref{firstauthor}}
\author[address]{Mingde Qin\corref{firstauthor}}
\author[address]{Chi Chen}
\author[address]{Weike Ye}
\author[address]{Xiangguo Li}
\author[address]{Jian Luo\corref{correspondingauthor}}
\author[address]{Shyue Ping Ong\corref{correspondingauthor}}

\cortext[firstauthor]{These authors contribute equally to this work}
\cortext[correspondingauthor]{Correspondences: jluo@ucsd.edu, ongsp@eng.ucsd.edu}
\address[address]{Department of NanoEngineering, University of California San Diego, 9500 Gilman Dr, Mail Code 0448, La Jolla, CA 92093-0448, United States}

\end{frontmatter}

\begin{figure}[H]
    \centering
    \includegraphics[width=0.8\textwidth]{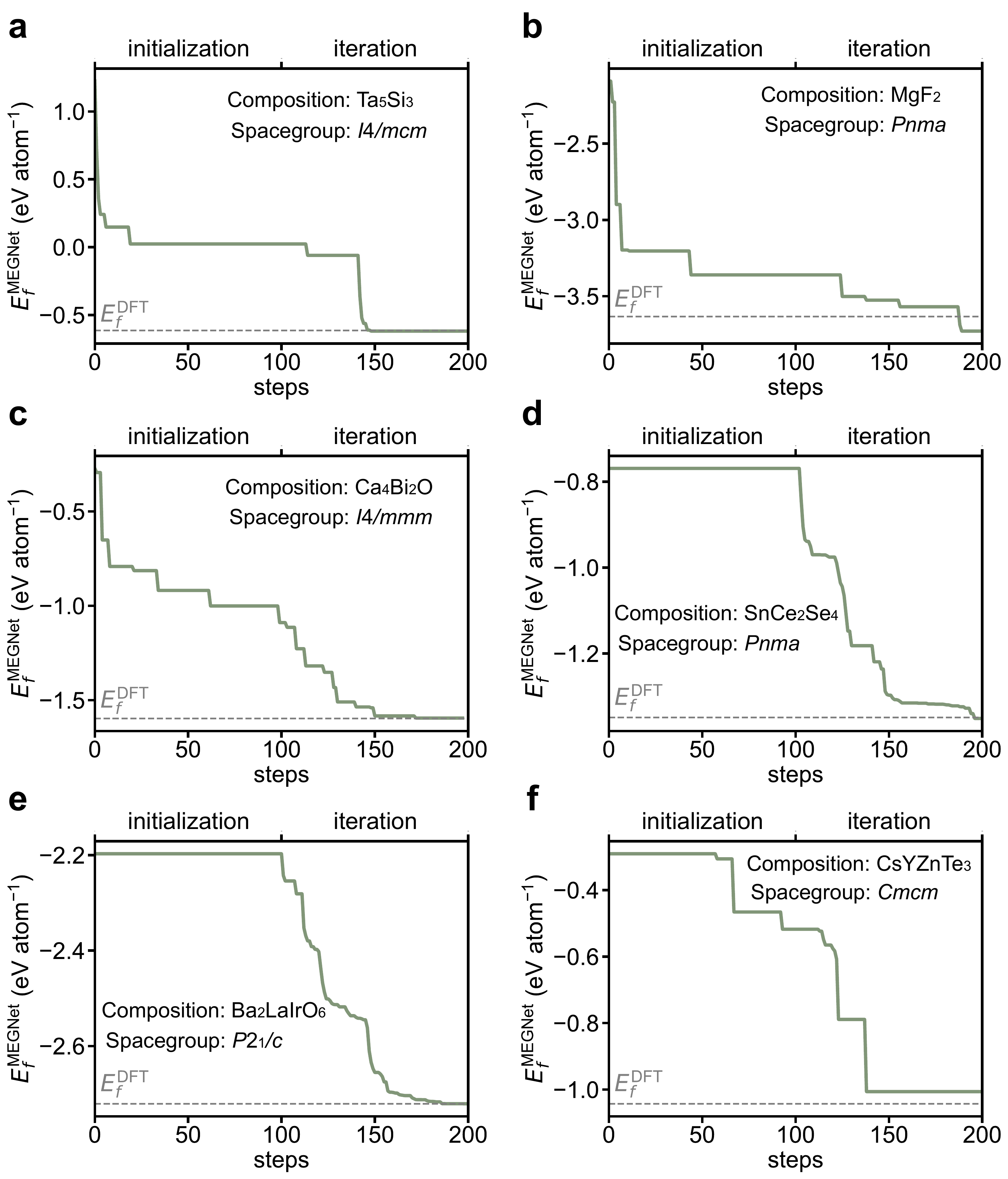}
    \caption{\textbf{Convergence of the BOWSR algorithm using the MEGNet energy model for six structures in the dataset of properties prediction.} These structures comprise two binary, two ternary, and two quaternary crystals obtained from elemental substitution in commonly occurring structure prototypes. The structure prototypes for \textbf{a, } \ce{Ta5Si3}, \textbf{b, } \ce{MgF2}, \textbf{c, } \ce{Ca4Bi2O}, \textbf{d, } \ce{SnCe2Se4}, \textbf{e, } \ce{Ba2LaIrO6}, \textbf{f, } \ce{CsYZnTe3} are \ce{Cr5B3} (ICSD\# 27124), \ce{Sr2Si} (ICSD\# 422), \ce{K2NiF4} (ICSD\# 15576), \ce{CaFe2O4} (ICSD\# 28177), \ce{La2ZnIrO6} (ICSD\# 75596), and \ce{KZrCuS3} (ICSD\# 80624), respectively. All structures were relaxed via the BOWSR algorithm using the default number of initialization samples (100) and iterations (100). }
    \label{fig:case}
\end{figure}

\pagebreak

\begin{figure}[H]
    \centering
    \includegraphics[width=1\textwidth]{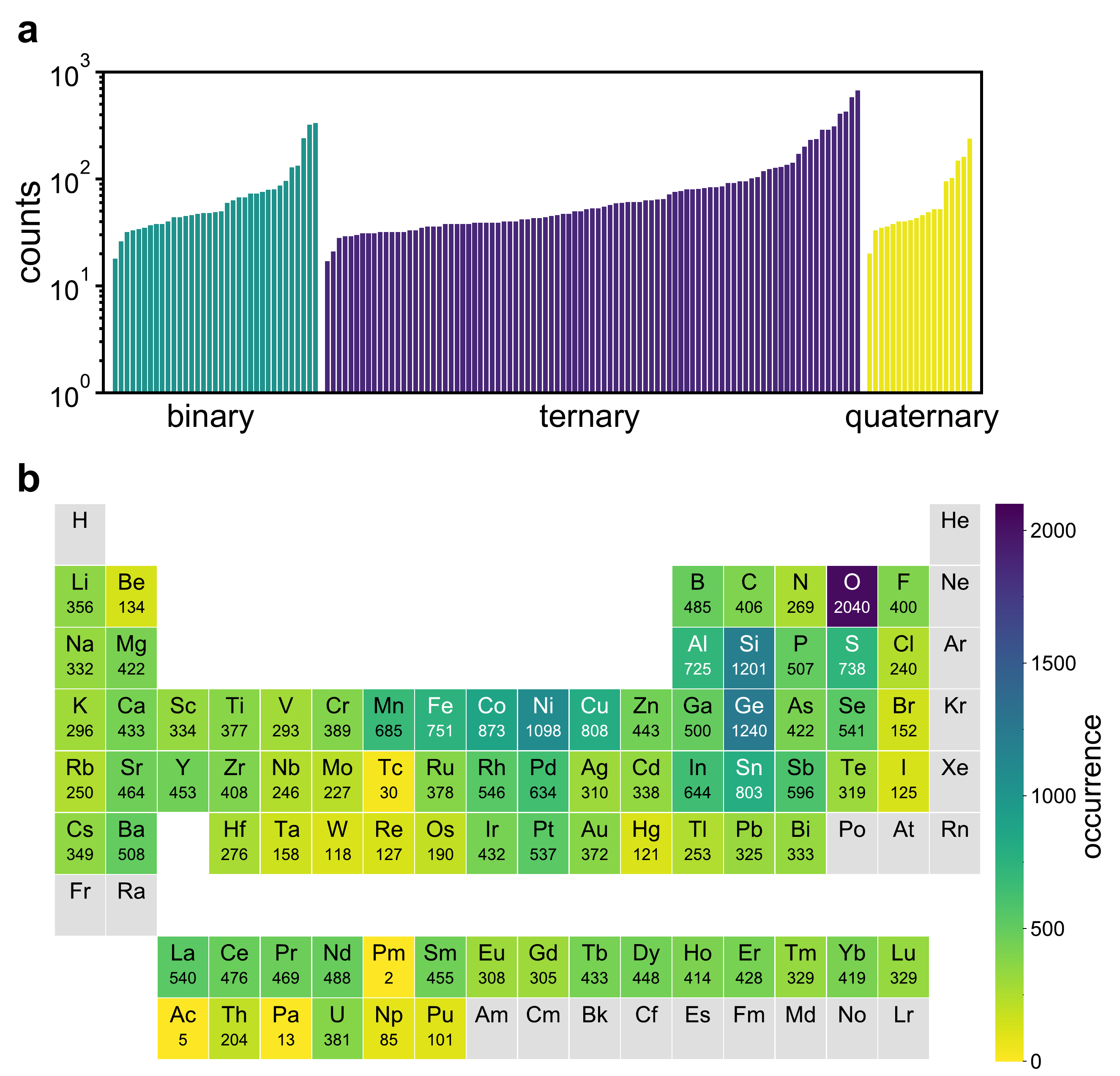}
    \caption{\textbf{Statistical distribution of dataset used for property predictions.} \textbf{a,} The distribution of 35 binary, 91 ternary, and 18 quaternary commonly occurring structure prototypes in the dataset. Each bar represents one structure prototype and there are at least 30 unique compositions for each structure prototype. \textbf{b,} Frequency of each element occurring in the dataset. Elements are color-coded according to the number of occurrences. Oxygen is the most common element. The relatively high frequencies of the transition metal elements Fe, Co, Ni, Cu can be attributed to the commonly occurring intermetallic structure prototypes. The three most commonly occurring structure prototypes are the ternary intermetallic (\ce{ThCr2Si2},\cite{banCrystalStructureTernary1965} \ce{TiNiSi},\cite{shoemakerTernaryAlloyPbCl2type1965} and \ce{ZrNiAl}\cite{dwightTernaryCompoundsFe2P}) and have 633, 619, and 484 compositions, respectively.
    }
    \label{fig:elements_statistic}
\end{figure}

\pagebreak

\begin{figure}[H]
    \centering
    \includegraphics[width=0.8\textwidth]{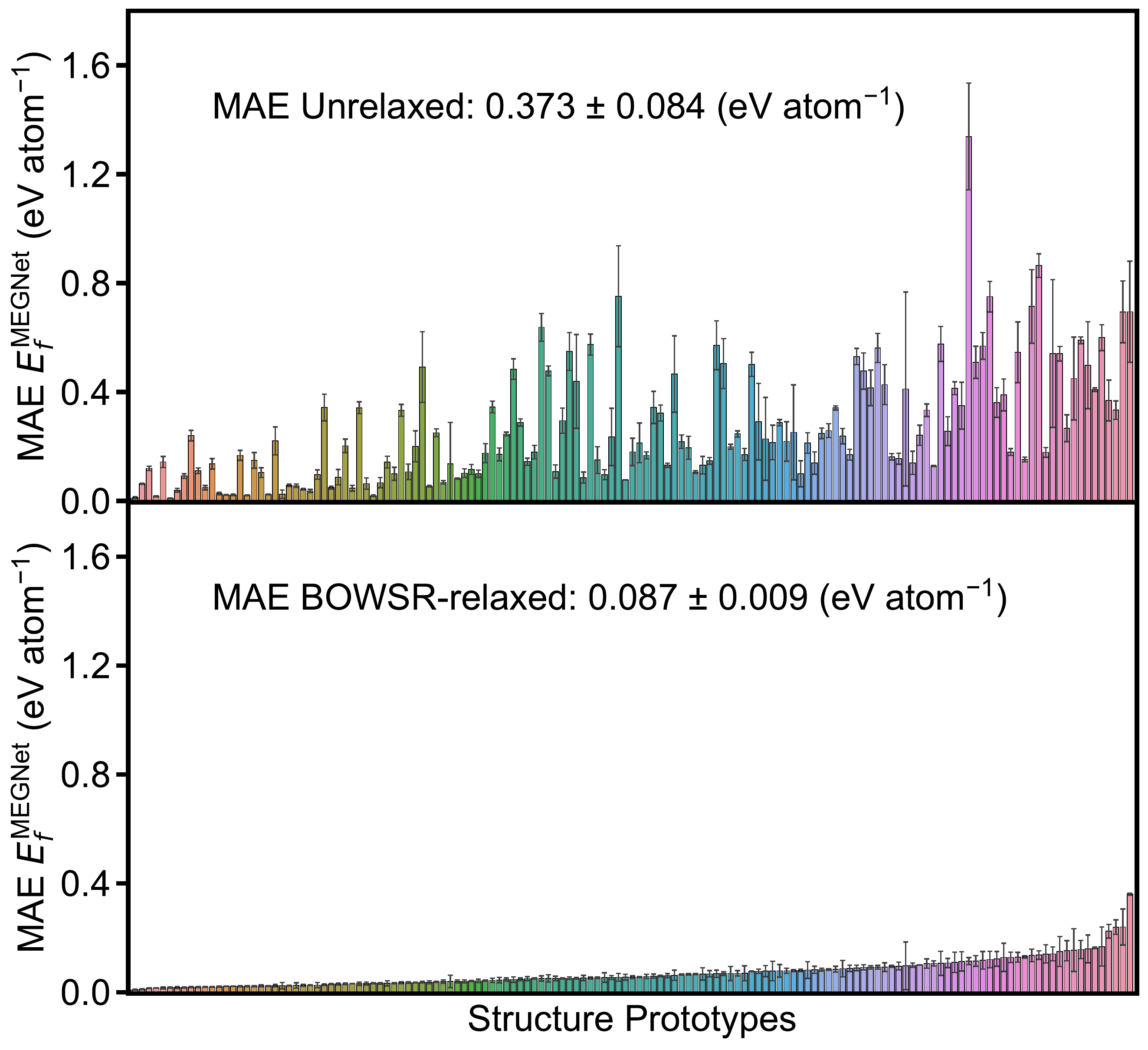}
    \caption{\textbf{Mean absolute errors (MAEs) of MEGNet prediction compared to DFT ground state calculations in formation energies using unrelaxed and BOWSR-relaxed structures grouped by structure prototypes.} To test the sensitivity of the BOWSR algorithm to the initial structures, within each structure prototype, we selected four parent structures with different lattice parameters for elemental substitution to obtain the unrelaxed structures (i.e., same composition with different lattice parameters). The mean and standard deviation in the MAEs of the MEGNet formation energy prediction with respect to DFT-computed values for each prototype are plotted in the ascending order of mean MAE for the BOWSR-relaxed structures. While the unrelaxed structures obtained from elemental substitution have large MAEs, relaxation via the BOWSR algorithm consistently yields structures with much lower and less noisy MAEs in the formation energies.}
    \label{fig:different_crystals}
\end{figure}

\begin{figure}[H]
    \centering
    \includegraphics[width=1\textwidth]{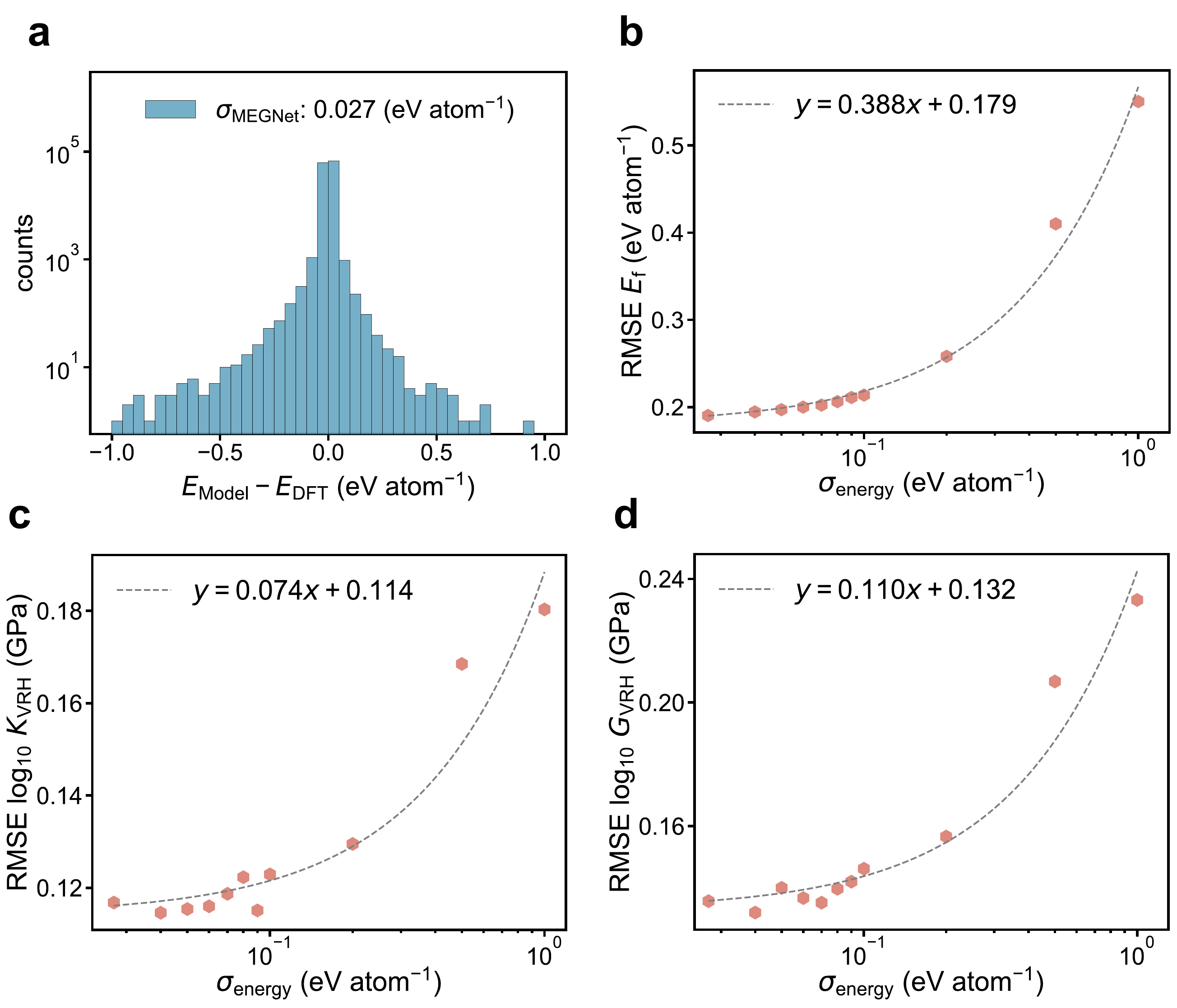}
    \caption{\textbf{Sensitivity of the BOWSR algorithm to the accuracy of the energy model.} \textbf{a, } Error distributions of the MEGNet formation energy model on the training and test data. Both the standard deviation ($\sigma$) and root mean squre error (RMSE) are 27 meV $\rm atom^{-1}$. Varying amounts of Gaussian noise are added to the MEGNet formation energy prediction during the BOWSR relaxation process. The error of the energy model $\sigma_{\rm energy}$ is then given by $\sqrt{\sigma_{\mathrm{MEGNet}}^2 + \sigma_{\mathrm{noise}}^2}$, where $\sigma_{\mathrm{noise}}$ is the standard deviation of the added noise. The RMSEs of the MEGNet-predicted \textbf{b, } formation energy, \textbf{c, } bulk modulus, and \textbf{d, } shear modulus for the BOWSR-relaxed structures are plotted against the error in the energy model. In all cases, linear correlations are observed between the RMSE of the MEGNet prediction and the error of the energy model, and reasonably low RMSEs in prediction are obtained when $\sigma_{\rm energy}< 0.1$ eV atom$^{-1}$.}
    \label{fig:sensitivity}
\end{figure}

\pagebreak

\begin{figure}[H]
    \centering
    \includegraphics[width=1\textwidth]{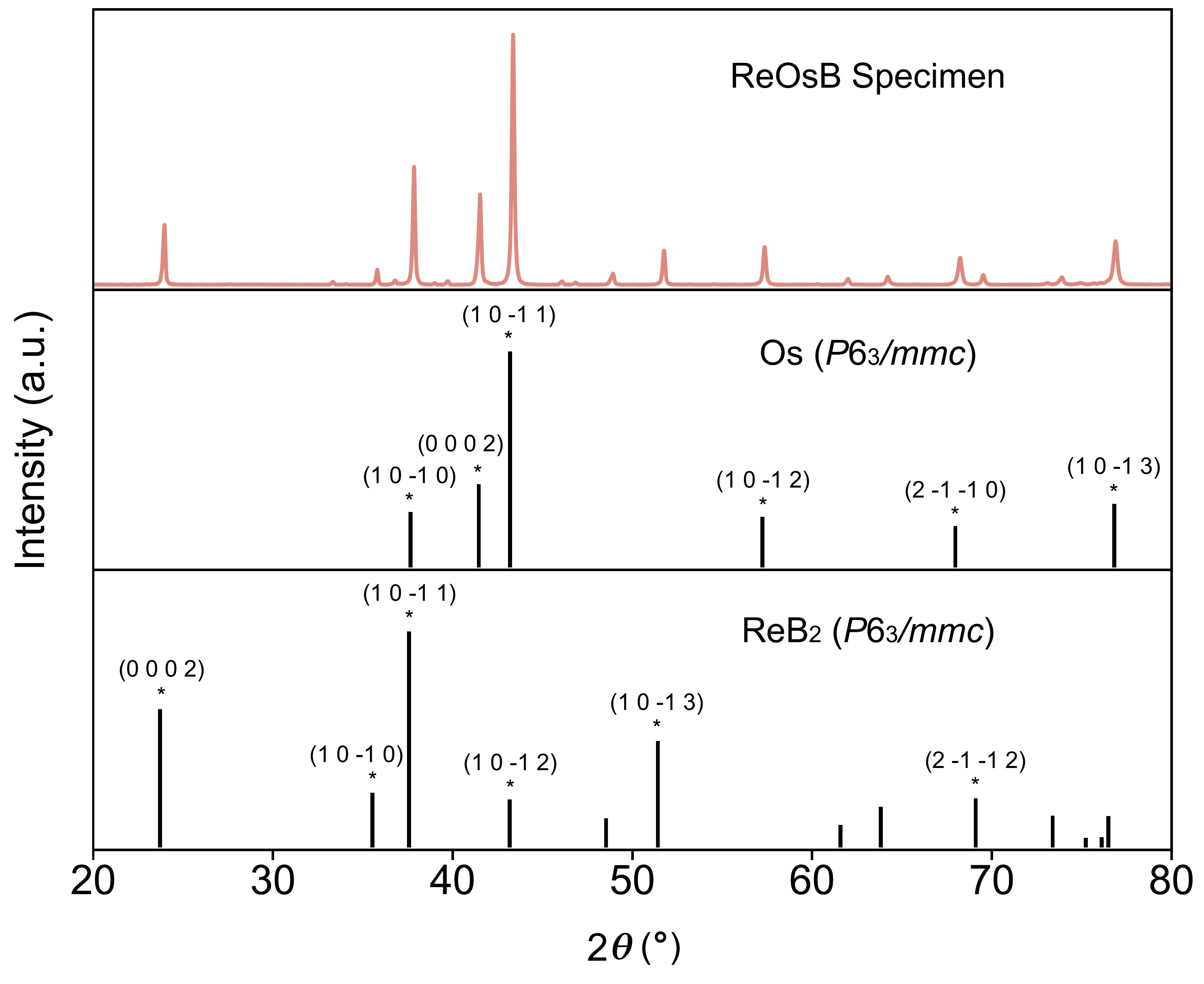}
    \caption{\textbf{Measured XRD patterns of the \ce{ReOsB} specimen.} The specimen exhibits two major phases: Os ($P6_{3}/mmc$) and \ce{ReB2} ($P6_{3}/mmc$). The major peaks are indexed for reference.}
    \label{fig:ReOsB}
\end{figure}

\begin{figure}[H]
    \centering
    \includegraphics[width=1\textwidth]{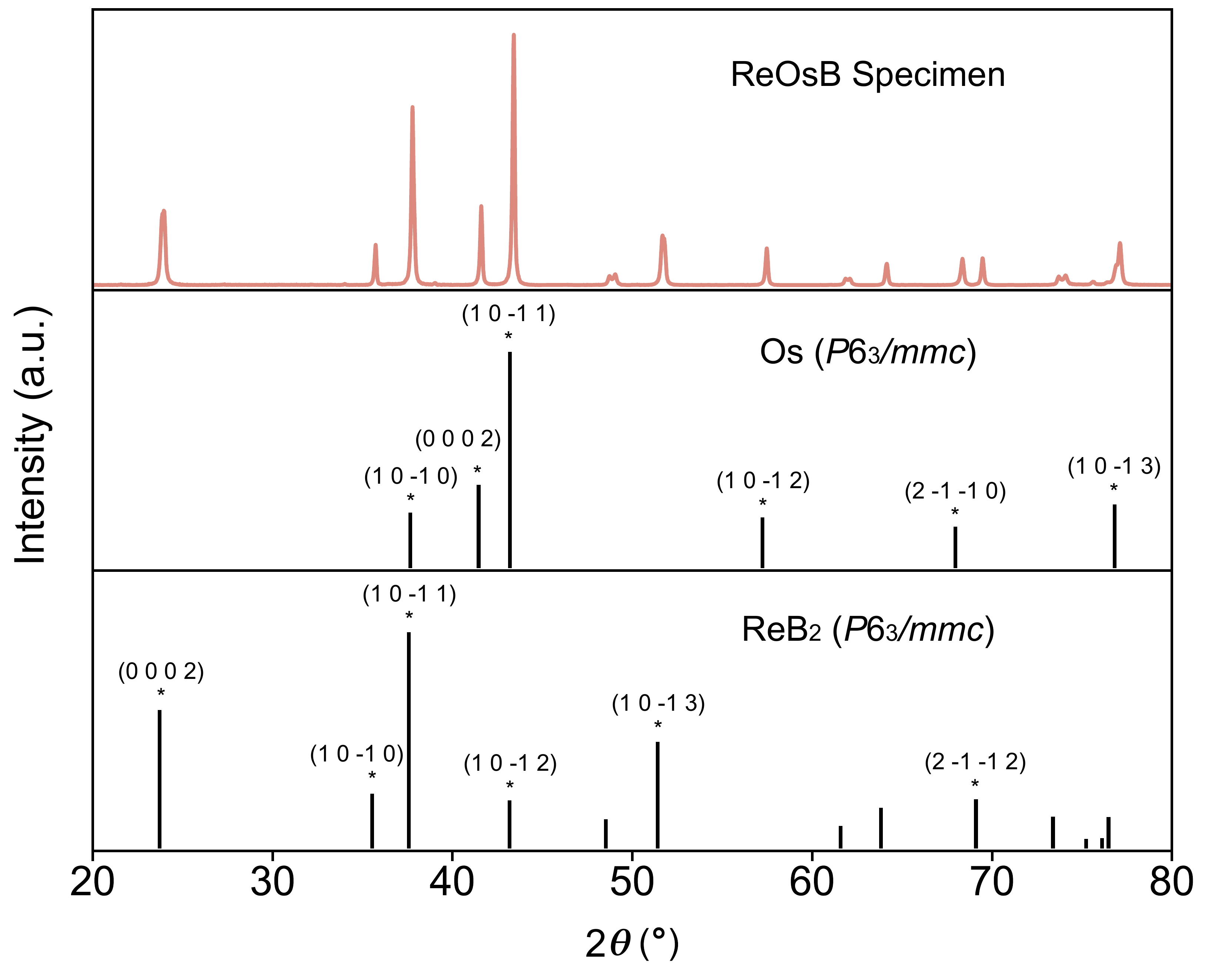}
    \caption{\textbf{Measured XRD patterns of the \ce{ReOsB2} specimen.} The specimen exhibits two major phases: Os ($P6_{3}/mmc$) and \ce{ReB2} ($P6_{3}/mmc$). The major peaks are indexed for reference.}
    \label{fig:ReOsB2}
\end{figure}

\begin{figure}[H]
    \centering
    \includegraphics[width=1\textwidth]{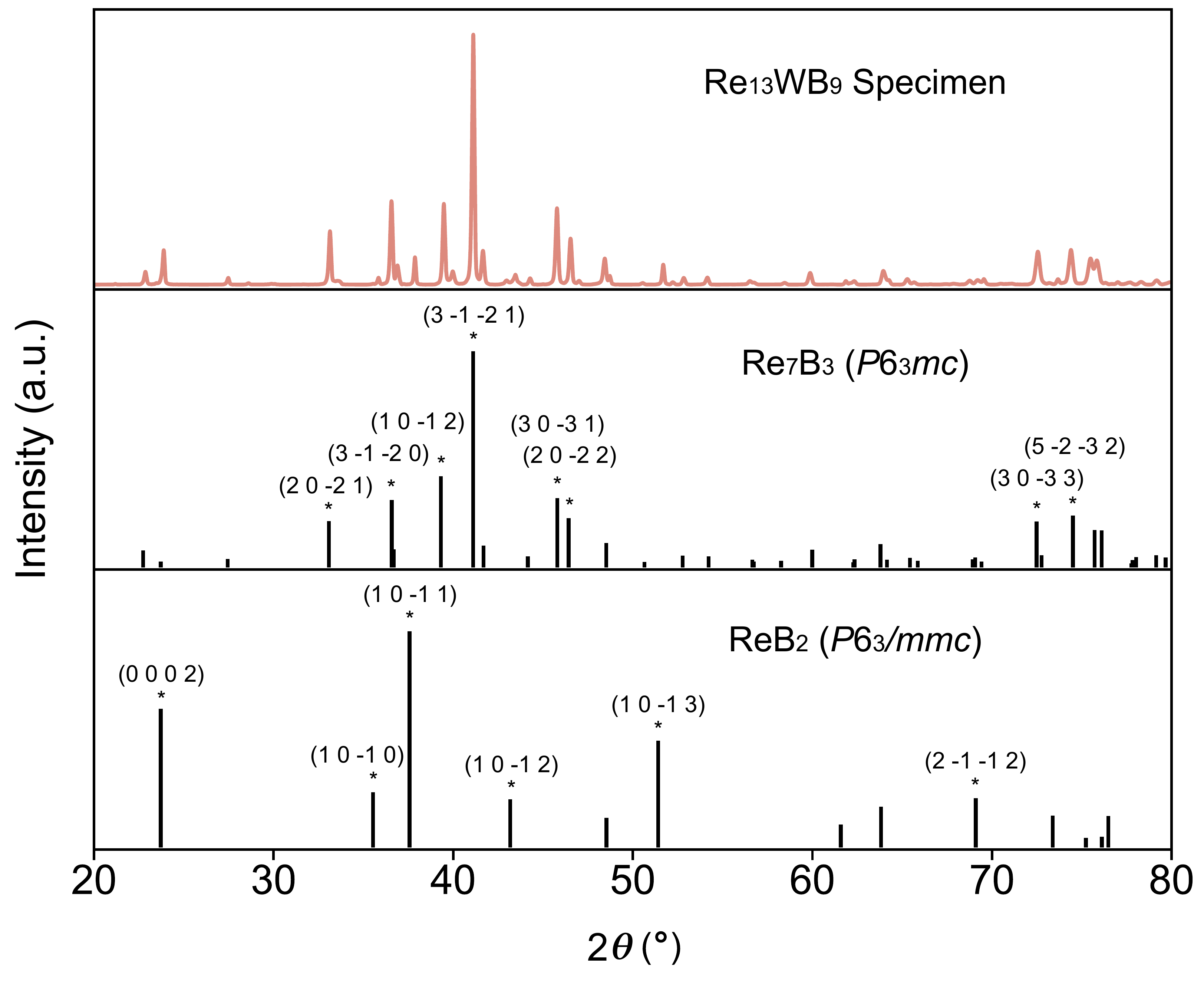}
    \caption{\textbf{Measured XRD patterns of the \ce{Re13WB9} specimen.} The specimen exhibits the primary phase of \ce{Re7B3} ($P6_{3}mc$) and the secondary phase of \ce{ReB2} ($P6_{3}/mmc$). The major peaks are indexed for reference.}
    \label{fig:Re13WB9}
\end{figure}

\begin{figure}[H]
    \centering
    \includegraphics[width=1\textwidth]{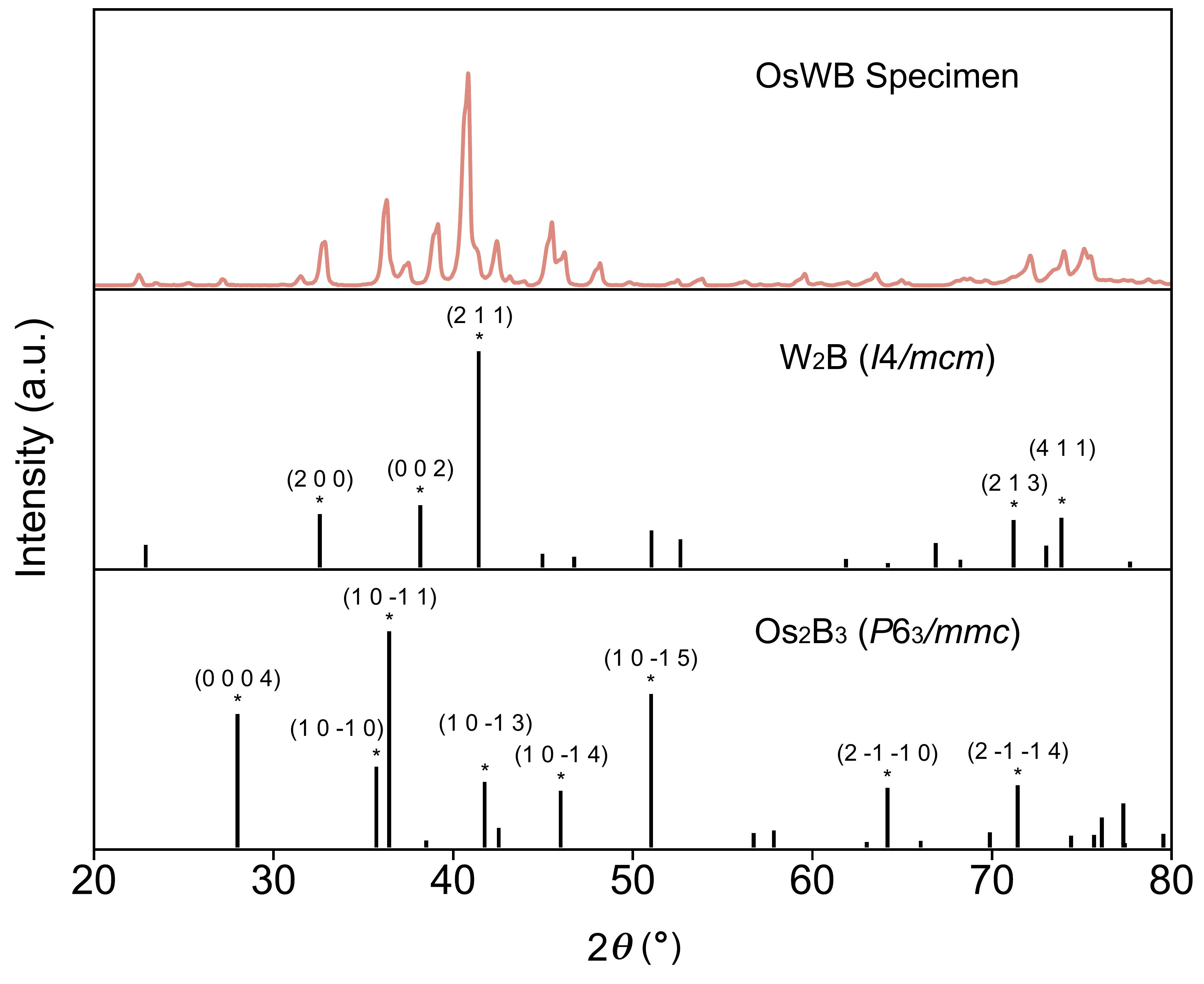}
    \caption{\textbf{Measured XRD patterns of the \ce{OsWB} specimen.} The specimen exhibits multiple phases including the \ce{W2B} ($I4/mcm$) and \ce{Os2B3} ($P6_{3}/mmc$). The major peaks are indexed for reference.}
    \label{fig:OsWB}
\end{figure}

\begin{figure}[H]
    \centering
    \includegraphics[width=1\textwidth]{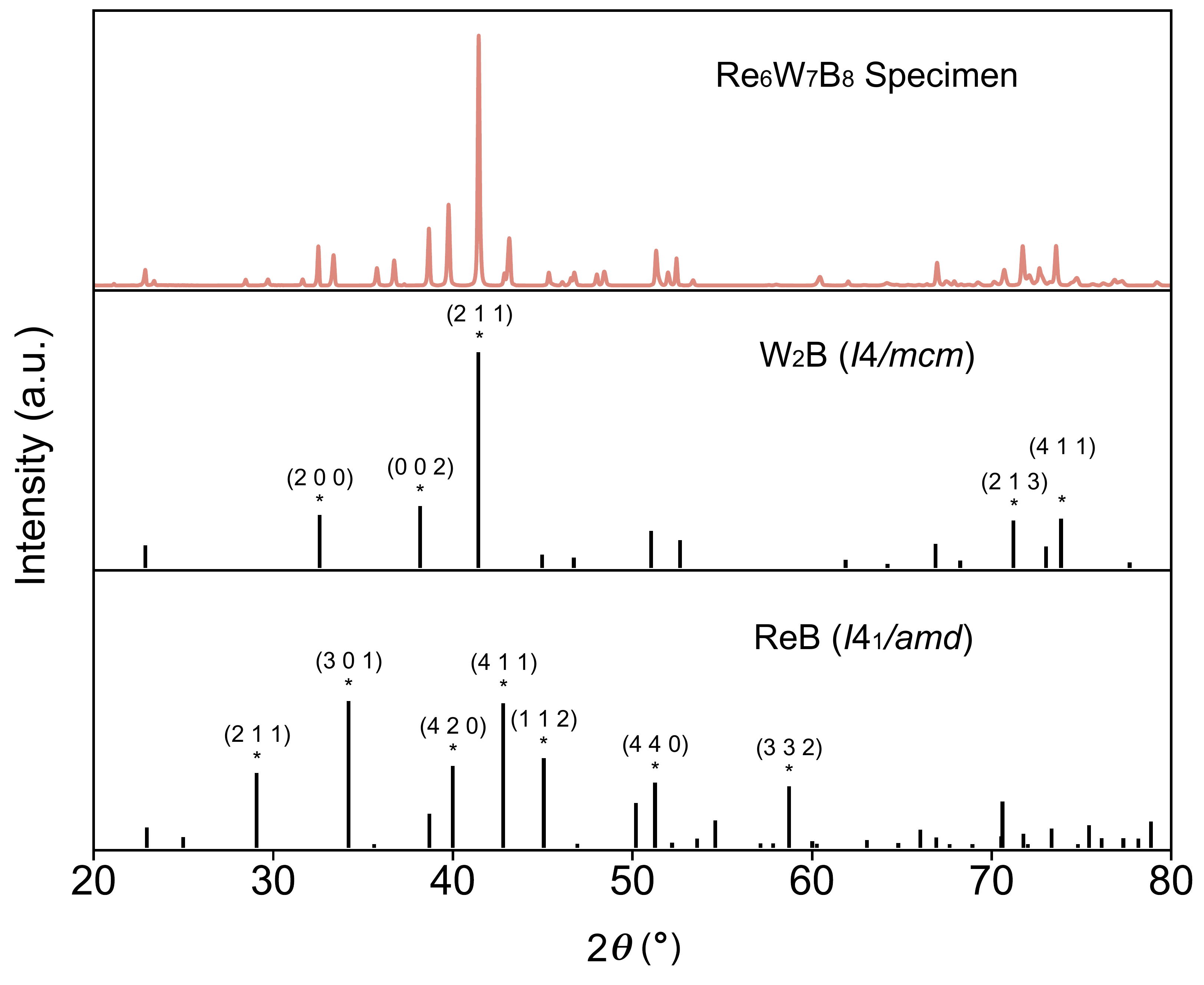}
    \caption{\textbf{Measured XRD patterns of the \ce{Re6W7B8} specimen.} The specimen exhibits the primary phase of \ce{W2B} ($I4/mcm$) and the secondary phase of \ce{ReB} ($I4_{1}/amd$). The major peaks are indexed for reference.}
    \label{fig:Re6W7B8}
\end{figure}

\begin{figure}[H]
    \centering
    \includegraphics[width=1\textwidth]{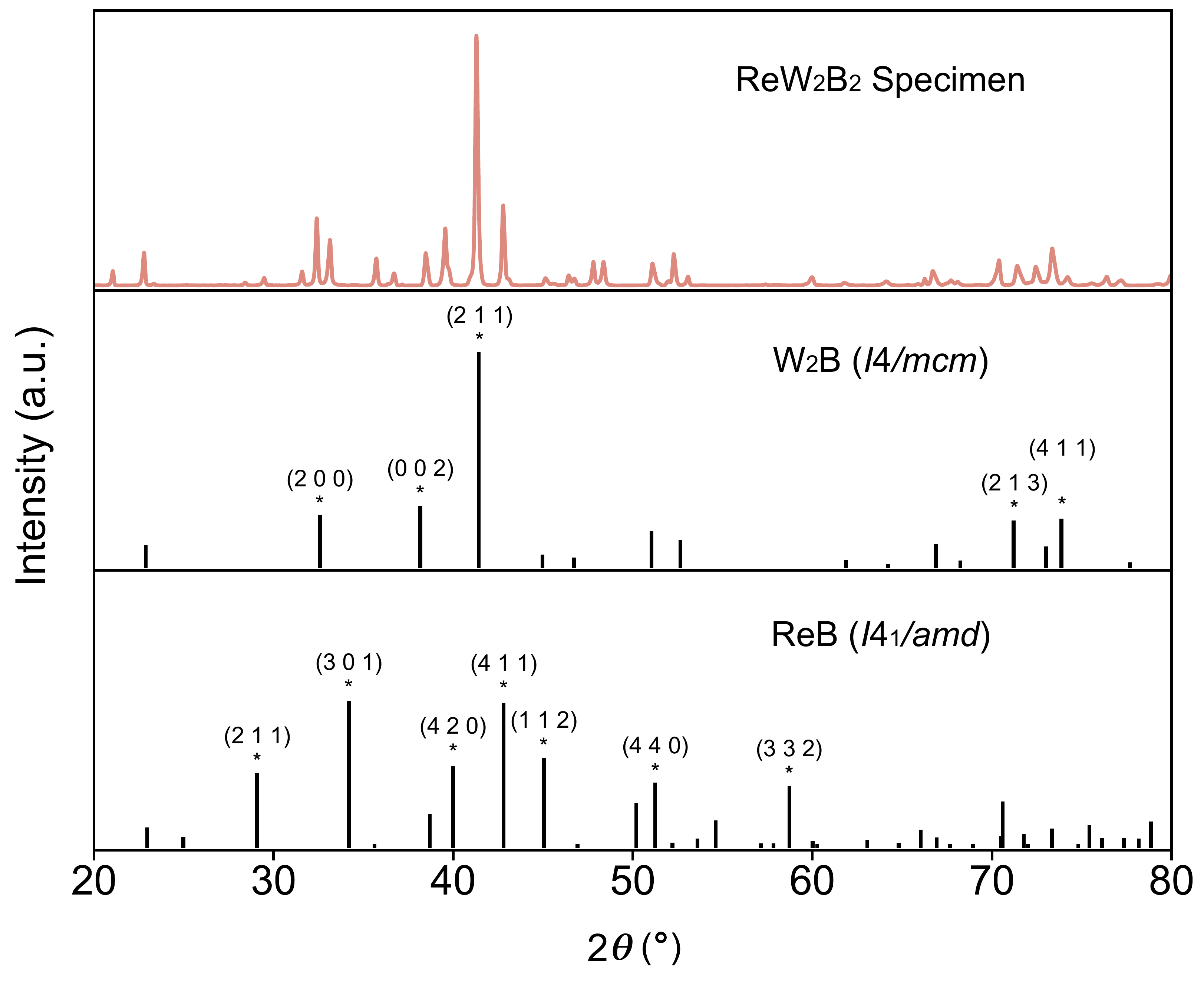}
    \caption{\textbf{Measured XRD patterns of the \ce{ReW2B2} specimen.} The specimen exhibits the primary phase of \ce{W2B} ($I4/mcm$) and the secondary phase of \ce{ReB} ($I4_{1}/amd$). The major peaks are indexed for reference.}
    \label{fig:ReW2B2}
\end{figure}

\pagebreak

\begin{figure}[H]
    \centering
    \includegraphics[width=1\textwidth]{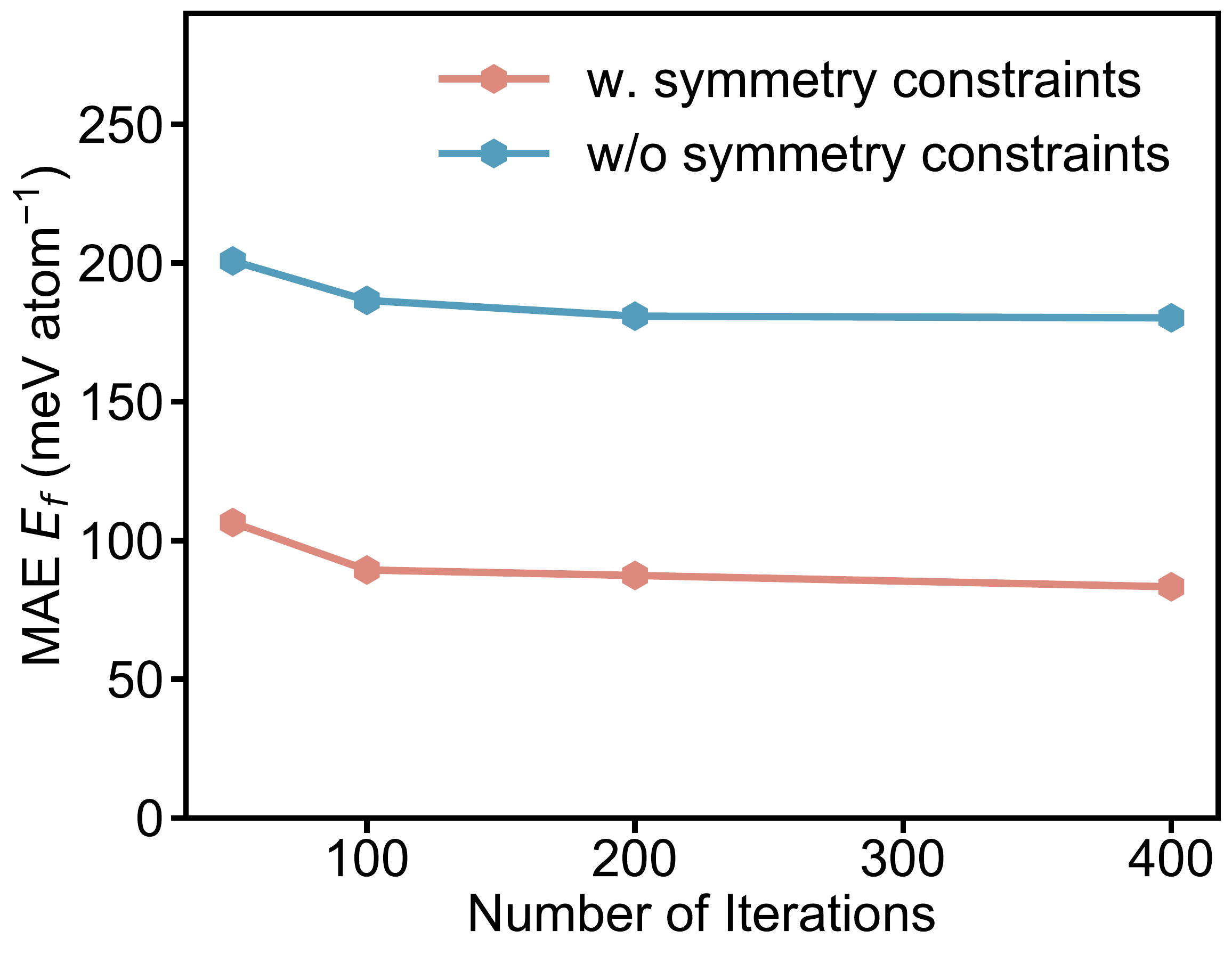}
    \caption{\textbf{Comparison of mean absolute errors (MAEs) of MEGNet predictions compared to DFT in formation energies of crystals relaxed using Bayesian optimization with and without symmetry constraints.} The MAEs are plotted with respect to the number of iterations.}
    \label{fig:convergence}
\end{figure}

\pagebreak

\begin{table}
\footnotesize
\begin{center}
\begin{tabular}{ ccccc }
\hline
\hline
& & BOWSR-relaxed & DFT-relaxed & Experimental\\
\hline
ReWB ($Pca2_1$) & $a$ (\AA) & 5.182 & 5.554 & 5.519\\
 & $b$ (\AA) & 4.864 & 5.537 & 5.506\\
 & $c$ (\AA) & 5.532 & 4.729 & 4.729\\
\hline
\ce{MoWC2} ($P6_3/mmc$) & $a$ (\AA) & 2.891 & 2.928 & 2.907\\
 & $c$ (\AA) & 11.580 & 11.448 & 11.372\\
\hline

\end{tabular}
\caption{\textbf{Comparison of the BOWSR-relaxed, DFT-relaxed and experimentally-measured lattice parameters for the discovered ultra-incompressible crystals, ReWB ($\textbf{\textit{Pca}}\textbf{2}_{\textbf{1}}$) and \ce{MoWC2} ($\textbf{\textit{P}}\textbf{6}_{\textbf{3}}/\textbf{\textit{mmc}}$).} Lattice angles are constrained to fixed values by virtue of their orthorhombic and hexagonal crystal systems.}
\label{table:lattice_parameters}
\end{center}
\end{table}

\bibliography{refs}